\documentclass[vecphys]{svmult}

\usepackage{graphicx}        
\usepackage{natbib}
\usepackage[bottom]{footmisc}

\begin{document}

\title*{The H$_{2}$ Amazing Life of NGC\,6881}
\author{Gerardo Ramos-Larios\inst{1}, Mart\'{\i}n A.\ Guerrero\inst{2}, 
\and Luis F.\ Miranda\inst{2}}
\institute{Instituto de Astronom\'{\i}a y Meteorolog\'{\i}a, \\
Av.\ Vallarta No.\ 2602, Col.\ Arcos Vallarta, C.P. 44130 Guadalajara, Jalisco, M\'exico. \\
\texttt{(gerardo@astro.iam.udg.mx)}
\and Instituto de Astrof\'{\i}sica de Andaluc\'{\i}a, IAA-CSIC, \\
C/ Camino Bajo de Hu\'etor 50, 18008 Granada, Spain. \\
 \texttt{(mar@iaa.es,lfm@iaa.es)}}
%
%
\maketitle

\begin{abstract}

The H$_{2}$ and optical (H$\alpha$, [N~{\sc ii}]) morphology of NGC\,6881 
are very different. 
Here we present a preliminary report of the analysis of new optical 
(H$\alpha$ and [N~{\sc ii}]) and near-IR (Br$\gamma$ and H$_2$) 
images and intermediate resolution $JHK$ spectra of this nebula.  
Our observations confirm the association of the H$_2$ bipolar lobes 
to NGC\,6881 and reveal that H$_2$ is predominantly shock excited 
in this nebula.  
We conclude that NGC\,6881 has multiple bipolar lobes that formed 
at different phases of the nebular evolution and that the collimation 
conditions or even the collimating agent changed from one ejection to 
the other.

\end{abstract}

\section{Introduction}
\label{sec:1}

Bipolar planetary nebulae (PNe) with multiple bipolar lobes have begun 
to be common.  
These objects are of great interest as they require multiple bipolar 
ejections with changes in the ejection direction in many cases.  
NGC\,6881 displays in the optical a quadrupolar morphology consisting 
of two pairs of highly collimated bipolar lobes aligned along different 
directions \citep{GM98}.  
There is also evidence that the axis of the central ring or torus has 
precessed in the last stages of PN formation \citep{KS05}.  
The distribution of the emission of the molecular hydrogen is very 
different from this of the ionized material \citep{Getal00}.  
Molecular hydrogen emission in NGC\,6881 is detected mainly in the 
equatorial region and in wide hourglass bipolar lobes that are much 
more extended than the ionized bipolar lobes.

The H$_2$ bipolar lobes of NGC\,6881 may represent an early bipolar 
ejection that took place before the formation of the two pairs of 
ionized bipolar lobes.  
To investigate in detail the spatial distribution of molecular hydrogen 
and ionized material within NGC\,6881 and to determine the prevalent 
excitation mechanism of the H$_{2}$ emission, we have obtained new near-IR 
Br$\gamma$ and H$_{2}$ (1,0) S(1) images, optical H$\alpha$ and 
[N~{\sc ii}] images, and intermediate resolution $JHK$ spectra of NGC\,6881.  
Preliminary results of our analysis are presented here.  
A more detailed analysis will be presented elsewhere (Ramos-Larios, 
Guerrero, \& Miranda, in prep.).


\section{Observations}
\label{sec:2}

\begin{figure}
\centering
\includegraphics[height=10.4cm]{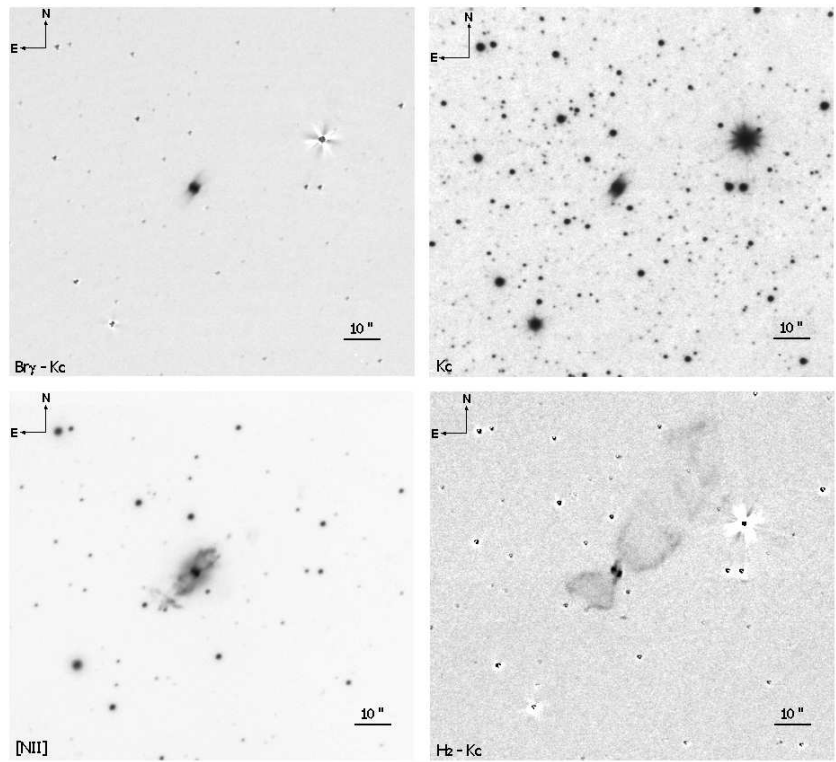}
\caption{
Images of NGC\,6881 in the Br$\gamma$ {\it (top-left)}, 
$K$c {\it (top-right)}, [N~{\sc ii}] {\it (bottom-left)}, 
and H$_2$ {\it (bottom-right)} narrowband filters.
}
\label{fig:1}      
\end{figure}

H$\alpha$ and [N~{\sc ii}] narrow-band images of NGC\,6881 were obtained 
with ALFOSC at the 2.5-m \emph{NOT} with total exposure times of 900 s 
and 1,800 s, respectively.  
H$_2$ (1,0) S(1), Br$\gamma$, and $K$c (continuum at $\lambda$ =2.270 $\mu$m)  
narrow-band images were obtained with LIRIS at the 4.2-m \emph{WHT}.  
Total exposure times were 1,000 s for Br$\gamma$ and $K$c, and 900 s for 
H$_2$.  
The images are presented in Figure~\ref{fig:1}.

Intermediate resolution $JHK$ longslit spectroscopy was obtained using 
NICS at the 3.5-m \emph{TNG}.  
The longslit was placed along the central star at PA's 113$^{\circ}$ 
and 137$^{\circ}$.  
At PA 113$^{\circ}$, spectra were obtained using the JH and $K_{\rm B}$ 
grisms with exposure times 1,800 s and 3,000 s, respectively. 
At PA 137$^{\circ}$, only the $K_{\rm B}$ grism was used with exposure 
time 1,200 s.

\begin{figure}
\centering
\includegraphics[height=8.0cm]{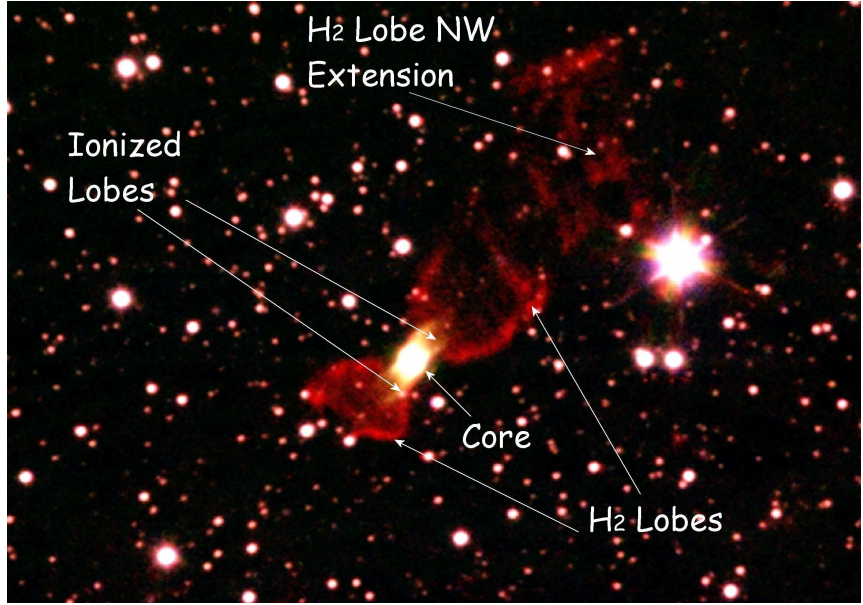}
\caption{
\emph{WHT} LIRIS composite three-color picture of NGC\,6881 in the filters 
of H$_2$ (red), Br$\gamma$ (green), and $K$c (blue).  
The picture is overlaid with regions of interest.
}
\label{fig:2}      
\end{figure}

\begin{figure}
\centering
\includegraphics[height=6.7cm]{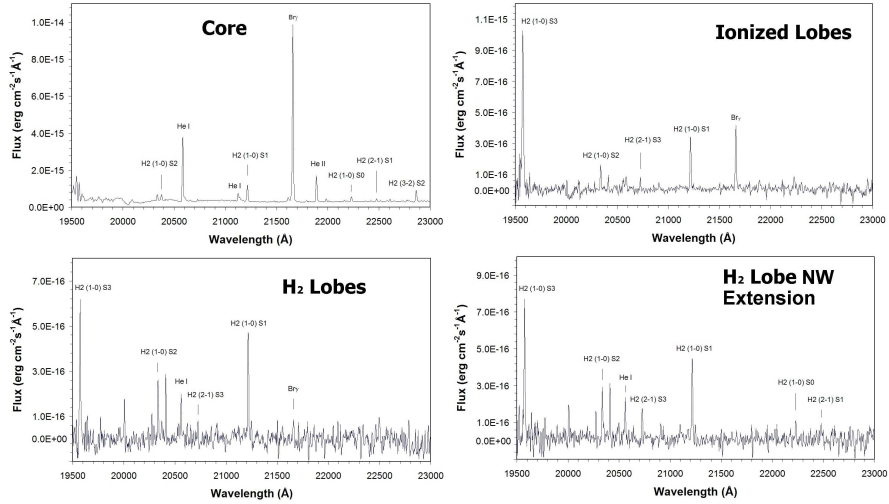}
\caption{
K$_B$ spectra along PA=137$^{\circ}$ of the different regions of 
NGC\,6881 described in the text and overlaid in Fig.~\ref{fig:2}: 
core, inner ionized lobes, H$_2$ lobes, and extension of the 
Northwest H$_2$ lobe.  
Some lines are overlaid on the spectra.}
\label{fig:3}       
\end{figure}

The new H$_2$, Br$\gamma$, and [N~{\sc ii}] images of NGC\,6881 
confirm the notable differences in the spatial distributions of 
ionized material and molecular hydrogen within this nebula 
\citep{Getal00}. 
Moreover, the better spatial resolution, depth, and accurate background 
subtraction in the H$_2$ image presented here and the invaluable 
information contained in the near-IR spectra have allowed us to obtain 
a more detailed view of the distribution and excitation of the molecular 
hydrogen in NGC\,6881.  
Both the images and spectra show that the outermost regions are 
H$_2$-dominated, while the central region and ionized bipolar 
lobes are emitting predominantly through lines of ionized species.  
This is clearly illustrated in the composite picture shown in 
Figure~\ref{fig:2}.  
Guided by this picture, we have defined four different regions of 
interest: the central region or core, the ionized bipolar lobes, 
the H$_2$ hourglass bipolar lobes, and the extension of the Northwest 
H$_2$ lobe.  
Individual spectra of these regions are shown in Figure~\ref{fig:3}, 
and the line identifications and fluxes are listed in Table~\ref{tab:1}.

The central region of NGC\,6881 shows the spatial distribution expected 
for a ring of molecular material surrounding the innermost ionized region. 
He~{\sc ii} emission is confined to this central region.  
The ionized lobes show both prominent Br$\gamma$ and H$_2$ emission, 
although their spectrum is probably contaminated by this of the 
H$_2$-dominated bipolar lobes.  
The H$_2$ lobes share the same orientation than the ionized lobes, but 
are less collimated and display a distinct hourglass shape.  
The Northern H$_2$ lobe extends 5 times farther than the ionized 
lobes, but the Southern H$_2$ lobe has a reduced extension. 
This asymmetry suggests the interaction of the nebula with an 
inhomogeneous interstellar medium. 
Intriguingly, the edge of the Southern H$_2$ lobe is coincident with 
the loop of ionized material seen in the [N~{\sc ii}] image, thus 
suggesting that it is not a precessing collimated outflow, as 
proposed by \cite{GM98}, but the outer edge of the Southern H$_2$ lobe.

The H$_2$ 1--0 S(1)/2--1 S(1) line ratio (see Tab.~\ref{tab:1}) 
derived for the different regions implies shock excitation even 
in the innermost regions of NGC\,6881.  
Excitation diagrams obtained using transitions from higher vibrational 
levels of the H$_2$ molecule confirm this result (Ramos-Larios et al., 
in prep.).

\section{Conclusions}

The very different morphologies of the ionized and molecular bipolar 
lobes of NGC\,6881 imply different bipolar ejections that most likely 
occurred at different times in the nebular evolution.  
While the direction of the bipolar ejection has only suffered 
small variations, especially at the very end of the nebular 
evolution \citep{KS05}, the degree of collimation has increased 
drastically.  
This suggests notable changes in the collimation conditions or even 
in the collimation mechanism during the formation of NGC\,6881.

\begin{table*}[!t]\centering
\caption{Line fluxes at PA 137$^{\circ}$ with $K_{\rm B}$ grism}
\label{tab:1}
\begin{tabular}{llccccc}
\hline
\multicolumn{2}{c}{} & \multicolumn{4}{c}{Regions} \\
\cline{3-6}
\multicolumn{1}{c}{$\lambda$} & 
\multicolumn{1}{c}{Line} & ~~Central~~ & ~~Ionized~~ & H$_2$ & H$_2$ Lobe  \\
\multicolumn{1}{c}{} & 
\multicolumn{1}{c}{Identification} & 
\multicolumn{1}{c}{} Region~~ & Lobes & Lobes & NW Extension \\
\multicolumn{1}{c}{[\AA]} & \multicolumn{1}{c}{} & 
\multicolumn{4}{c}{[~ergs~~cm$^{-2}$~~s$^{-1}$~~\AA$^{-1}$~]} \\
\hline
19446 & Br8 HI               & 6.3$\times$10$^{-14}$	&	$\dots$	&	$\dots$	&	$\dots$ \\
19549 & He I	             & 1.6$\times$10$^{-14}$	&	$\dots$	&	$\dots$	&	$\dots$ \\
19574 & H$_2$ (1,0) S(3)     & 1.2$\times$10$^{-14}$	&	1.7$\times$10$^{-14}$	&	1.1$\times$10$^{-14}$	&	9.2$\times$10$^{-15}$ \\
19703 & H$_2$ (8,6) O(2)~~~~ & 2.1$\times$10$^{-15}$	&	$\dots$	&	$\dots$	&	$\dots$ \\
20338 & H$_2$ (1,0) S(2)     & 4.2$\times$10$^{-15}$	&	2.1$\times$10$^{-15}$	&	3.8$\times$10$^{-15}$	&	4.6$\times$10$^{-15}$ \\
20377 & ?	             & 4.7$\times$10$^{-15}$	&	$\dots$	&	$\dots$	&	$\dots$ \\
20587 & He I	             & 4.7$\times$10$^{-14}$	&	1.3$\times$10$^{-15}$	&	4.2$\times$10$^{-16}$	&	2.4$\times$10$^{-15}$ \\
20732 & H$_2$ (2,1) S(3)	&	1.2$\times$10$^{-15}$	&	8.1$\times$10$^{-16}$	&	1.3$\times$10$^{-15}$	&	2.8$\times$10$^{-15}$ \\
20763 & ?	&	8.5$\times$10$^{-16}$	&	$\dots$	&	$\dots$	&	$\dots$ \\
21126 & He I	&	5.1$\times$10$^{-15}$	&	$\dots$	&	$\dots$	&	$\dots$\\
21138 & He I	&	2.2$\times$10$^{-15}$	&	$\dots$	&	$\dots$	&	$\dots$ \\
21218 & H$_2$ (1,0) S(1)	&	1.2$\times$10$^{-14}$	&	4.7$\times$10$^{-15}$	&	5.6$\times$10$^{-15}$	&	6.2$\times$10$^{-15}$ \\
21542 & H$_2$ (2,1) S(2)	&	4.1$\times$10$^{-16}$	&	$\dots$	&	$\dots$	&	$\dots$ \\
21614 & He I ??	&	2.9$\times$10$^{-15}$	&	$\dots$	&	$\dots$	&	$\dots$\\
21658 & Br7 HI (Brg)	&	1.1$\times$10$^{-13}$	&	5.5$\times$10$^{-15}$	&	1.3$\times$10$^{-15}$	&	$\dots$ \\
21793 & Br I	&	1.1$\times$10$^{-15}$	&	$\dots$	&	$\dots$	&	$\dots$ \\
21887 & He II	&	1.7$\times$10$^{-14}$	&	8.3$\times$10$^{-16}$	&	$\dots$	&	$\dots$ \\
21985 & ?	&	2.1$\times$10$^{-15}$	&	$\dots$	&	$\dots$	&	$\dots$ \\
22235 & H$_2$ (1,0) S(0)	&	3.3$\times$10$^{-15}$	&	1.2$\times$10$^{-15}$	&	$\dots$	&	1.5$\times$10$^{-15}$ \\
22477 &	H$_2$ (2,1) S(1)	&	1.6$\times$10$^{-15}$	&	$\dots$	& $\dots$	&	9.6$\times$10$^{-16}$ \\
22872 &	Br I ??	&	7.1$\times$10$^{-15}$	&	$\dots$	&	$\dots$	&	$\dots$ \\
\hline
\end{tabular}
\vspace{0.4cm}
\end{table*}


\subsubsection{Acknowledgments}

This work is funded by grant PNAYA2005-01495 of the Spanish MEC.
GRL thanks the IAA for his hospitality and CONACyT (Mexico).  
We thank G.\ G\'omez and J.\ Acosta for taking the LIRIS images of NGC\,6881.
%
%
%
%




\begin{thebibliography}{99.}
%


\bibitem[Guerrero \& Manchado(1998)]{GM98} 
Guerrero, M. A., \& Manchado, A.\ 1998, ApJ, 279, 125
\bibitem[Guerrero et al.(2000)]{Getal00} 
Guerrero, M. A., Villaver, E., Manchado, A., Garc\'{\i}a-Lario, P., \& 
Prada, F.\ 2000, ApJS, 127, 125
\bibitem[Kwok \& Su(2005)]{KS05} 
Kwok, S., \& Su, K.Y.L.\ 2005, ApJ, 635, L49

\end{thebibliography}
\end{document}